\documentclass[twoside]{dis04}
\def\runauthor{A. Sabio~Vera}
\def\shorttitle{The Gluon Green's Function at Small x}
\begin{document}

\title{The Gluon Green's Function at Small x}

\author{Agust{\' \i}n Sabio Vera}

\address{II. Institut f{\"u}r Theoretische Physik, Universit{\"a}t Hamburg\\ Luruper Chaussee 149, 22761~Hamburg, Germany.\\
E-mail: sabio@mail.desy.de}

\maketitle

\abstracts{In this contribution a recently proposed iterative procedure is 
used to study the BFKL gluon Green's function at next--to--leading order. This 
is done in QCD and in N=4 supersymmetric Yang--Mills theory. The study 
includes an analysis of the evolution with energy and of angular dependences. 
A discussion of a novel resummation of running coupling terms in the QCD 
case is included.}

\section{Introduction}

The Balitsky--Fadin--Kuraev--Lipatov (BFKL)~\cite{FKL} equation resums 
a class of logarithms dominant in the Regge limit of scattering amplitudes 
where the centre of mass energy $\sqrt{s}$ is large and the momentum transfer 
$\sqrt{-t}$ is fixed. The energy dependence of the cross section is 
carried by the gluon Green's function (GGF), which describes the interaction 
among Reggeised gluons exchanged in the $t$--channel. The GGF carries 
a dependence on the transverse momenta of the exchanged gluons, 
$\vec{k}_{a,b}$,
 and evolves with energy following the BFKL equation, with the rapidity, Y, 
playing the role of a time variable. The equation can be written in terms
 of a Mellin transform in Y of the GGF, 
\begin{eqnarray}
f \left(\vec{k}_a,\vec{k}_b, {\rm Y}\right) 
&=& 
\int_{a-i \infty}^{a+i \infty} \frac{d\omega}{2 \pi i} ~ e^{\omega {\rm Y}} f_{\omega} 
\left(\vec{k}_a ,\vec{k}_b\right),
\label{Mellin}
\end{eqnarray}
and reads
\begin{eqnarray}
\omega f_\omega \left(\vec{k}_a,\vec{k}_b\right) &=& \delta^{(2)} 
\left(\vec{k}_a-\vec{k}_b\right) + \int d^{2}\vec{k}' ~
\mathcal{K}\left(\vec{k}_a,\vec{k}'\right)f_\omega \left(\vec{k}',\vec{k}_b 
\right).
\label{first}
\end{eqnarray}
The kernel, $\mathcal{K}$, is known at next--to-leading (NLL) accuracy where terms of the 
form $\left(\alpha_s {\rm Y} \right)^n$ and 
$\alpha_s \left(\alpha_s {\rm Y} \right)^n$ are 
resummed~\cite{Fadin:1998py,Ciafaloni:1998gs}. 
To solve the equation with no approximations, as shown in 
Ref.~\cite{Andersen:2003an},
 it is useful to use dimensional regularisation and treat the cancellation 
of infrared divergences without angular averaging the kernel. 
The solution can then be presented in the iterative form
\begin{eqnarray}
  f(\vec{k}_a ,\vec{k}_b, {\rm Y})  &=&
  e^{\omega_0 \left(\vec{k}_a,\lambda\right) {\rm Y}}
  \delta^{(2)} (\vec{k}_a - \vec{k}_b) \nonumber\\
  &&\hspace{-2.3cm}
+\sum_{n=1}^{\infty} \left\{\prod_{i=1}^{n} \int d^2 \vec{k}_i \int_0^{y_{i-1}}
  d y_i \left[\frac{\theta\left(\vec{k}_i^2 - \lambda^2\right)}{\pi \vec{k}_i^2} \, \xi \left(\vec{k}_i\right) 
\, +\widetilde{\mathcal{K}}_r \left(\vec{k}_a+\sum_{l=0}^{i-1}\vec{k}_l,
\vec{k}_a+\sum_{l=1}^{i}\vec{k}_l\right)\right] \right.\nonumber\\
&&\hspace{-1.6cm}\times \left. e^{
\omega_0\left(\vec{k}_a+\sum_{l=1}^{i-1}
  \vec{k}_l,\lambda\right) (y_{i-1}-y_i)}\right\}
e^{\omega_0\left(\vec{k}_a+\sum_{l=1}^{n}
  \vec{k}_l,\lambda\right) y_n} \delta^{(2)} \left(\sum_{l=1}^{n}\vec{k}_l 
+ \vec{k}_a - \vec{k}_b \right),
\label{TheSolution}
\end{eqnarray}
where $y_0 \equiv Y$. The functions $\omega_0$, $\xi$ and 
$\widetilde{\mathcal{K}}_r$ are described below when the behaviour 
of the solution is discussed.

\section{The GGF in the QCD case}

\begin{figure}[!thb]
\vspace*{5.1cm}
\begin{center}
\includegraphics{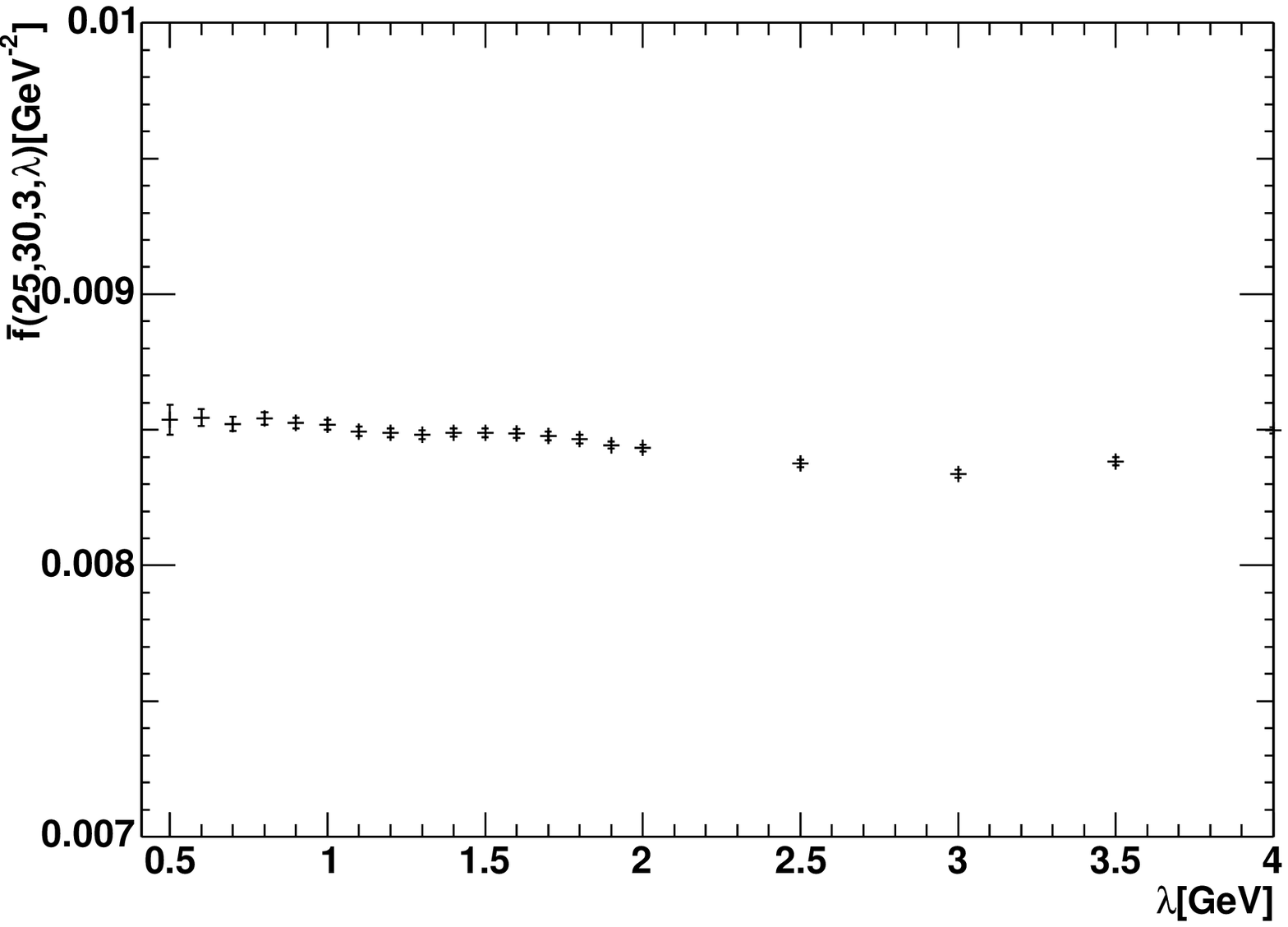}
\includegraphics{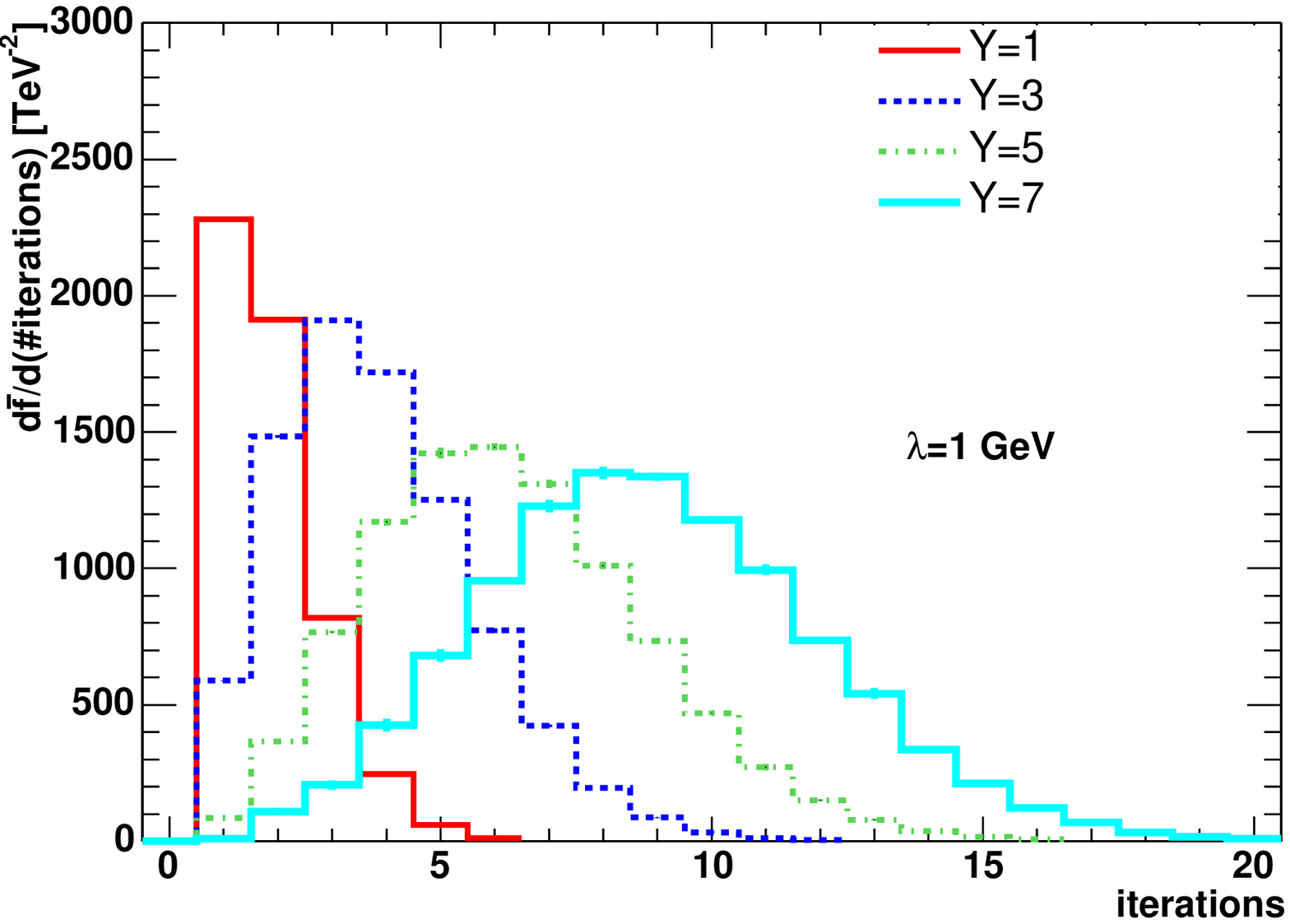}
\caption[*]{Left: $\lambda$ independence of the GGF. Right: Distribution 
on the number of iterations needed to calculate the GGF.}
\label{fig1}
\end{center}
\end{figure}

To regularise the infrared divergences a phase space slicing parameter 
$\lambda$ is introduced resulting in the following expressions:
\begin{eqnarray}
\omega_0 \left(\vec{q},\lambda\right) &\equiv&\lim_{\epsilon \to 0} \left( 2\, \omega^{(\epsilon)}\left(\vec{q}\right) + \int d^{2+2\epsilon}\vec{k} \,
\mathcal{K}_r^{(\epsilon)} \left(\vec{q},\vec{q}+\vec{k}\right) 
\theta \left(\lambda^2-\vec{k}^2\right) \right) \nonumber\\
&&\hspace{-2.4cm}= - \bar{\alpha}_s \left\{\ln{\frac{\vec{q}^2}{\lambda^2}}
+ \frac{\bar{\alpha}_s}{4}\left[\frac{\beta_0}{2 N_c}\ln{\frac{\vec{q}^2}{\lambda^2}}\ln{\frac{\mu^4}{\vec{q}^2 \lambda^2}}+\left(\frac{4}{3}-\frac{\pi^2}{3}+\frac{5}{3}\frac{\beta_0}{N_c}\right)\ln{\frac{\vec{q}^2}{\lambda^2}}-6 \zeta(3)\right]\right\},
\label{QCDtrajectory}
\end{eqnarray}
which will be the gluon Regge trajectory in this regularisation, and 
\begin{eqnarray}
\xi \left(\vec{q}\right) \equiv \bar{\alpha}_s +  
\frac{{\bar{\alpha}_s}^2}{4}\left(\frac{4}{3}-\frac{\pi^2}{3}+\frac{5}{3}\frac{\beta_0}{N_c}-\frac{\beta_0}{N_c}\ln{\frac{\vec{q}^2}{\mu^2}}\right),
\end{eqnarray}
which enters in the real emission integral. $\mu$ is the renormalisation 
scale in the $\overline{\rm MS}$ scheme. The remaining part of the kernel,  
$\widetilde{\mathcal{K}}_r$, can be found in Ref.~\cite{Andersen:2003an}. The 
solution as obtained from Eq.~(\ref{TheSolution}) is independent of $\lambda$ when this parameter is 
small compared to the external transverse scales entering the GGF. As an 
example in Fig.~\ref{fig1} (left) it is shown how the GGF for large external 
scales is independent of $\lambda$ for values of $\lambda$ well above 
$\Lambda_{\rm QCD}$, a consequence of the infrared finiteness of the calculation.
 For a given value of $\lambda$ convergence of the 
solution is achieved for a finite number of iterations as can be seen in 
Fig.~\ref{fig1} (right) where the GGF can be obtained as the area under the 
curves. The number of iterations of the kernel needed to 
obtain the solution increases with the available energy Y. These results have 
been published in Ref.~\cite{Andersen:2003wy}.

\begin{figure}[!thb]
\vspace*{5.0cm}
\begin{center}
\includegraphics{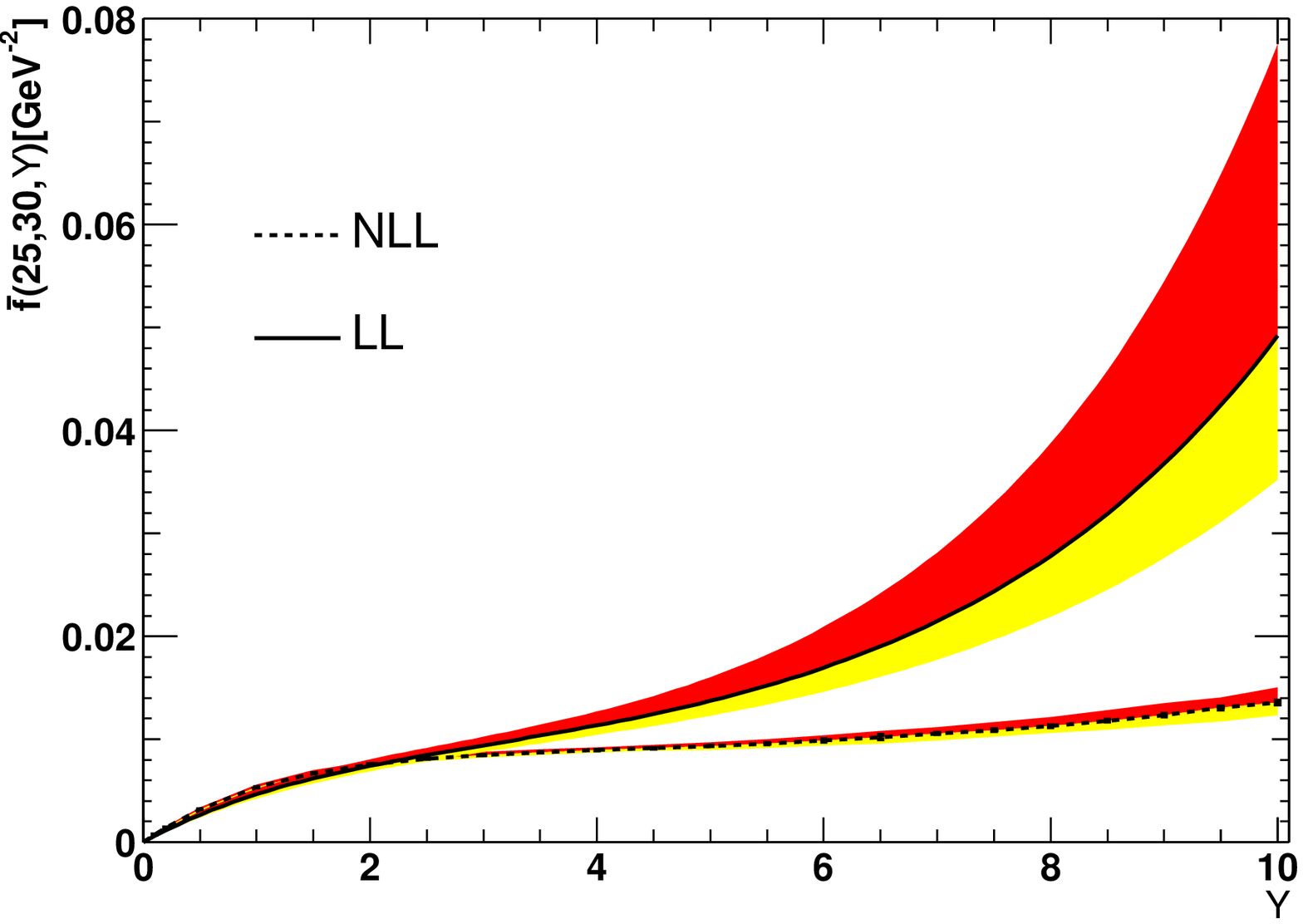}
\includegraphics{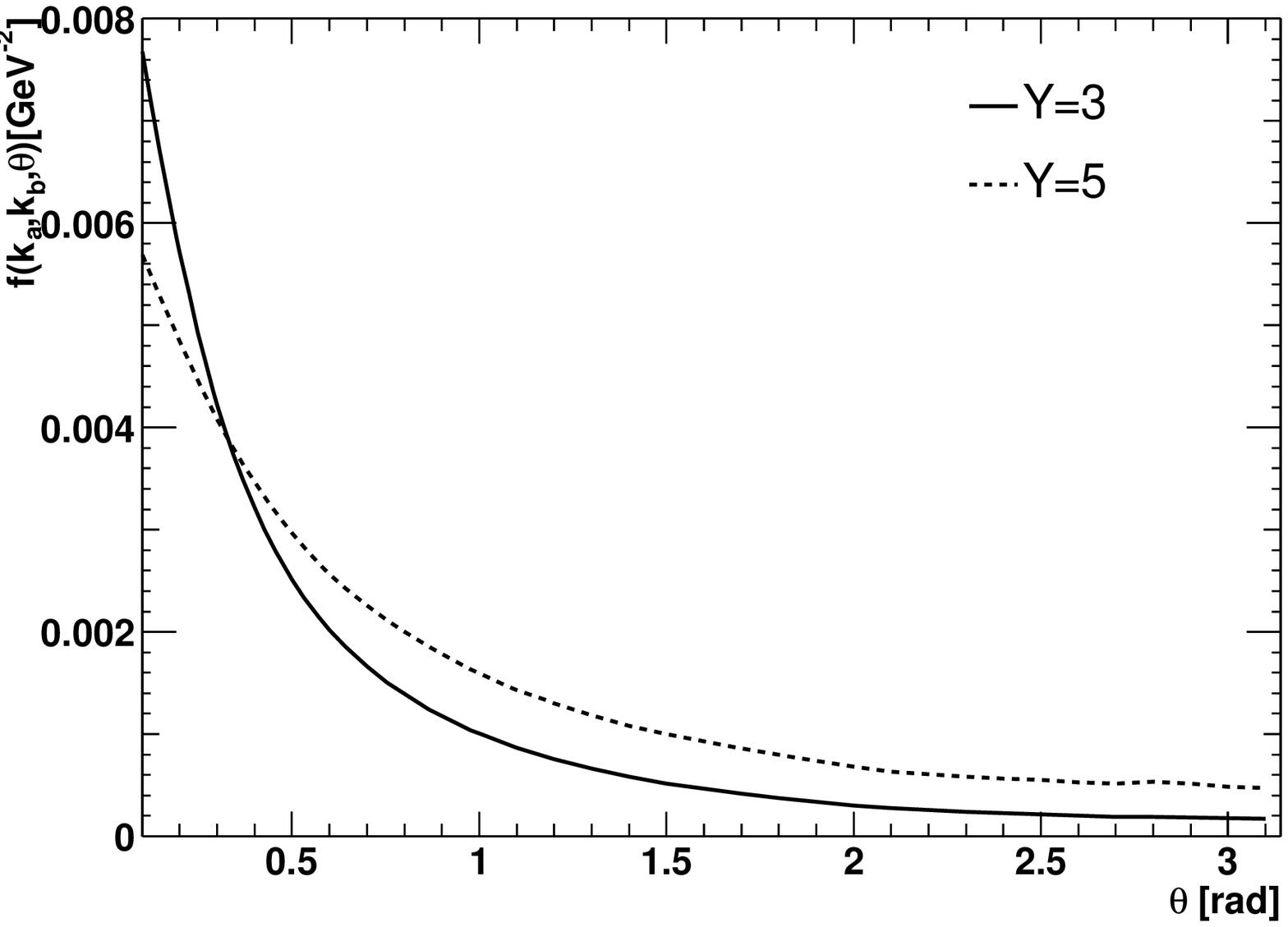}
\caption[*]{Left: Energy dependence of the GGF. Right: Angular dependence of 
the GGF.}
\label{fig2}
\end{center}
\end{figure}

In Fig.~\ref{fig2} (left) 
the growth with energy of the GGF is compared at LL to 
that at NLL. When higher order corrections are included the intercept 
diminishes. The bands correspond to the choices of renormalisation 
scale $\mu = k_b, k_b/2, 2 k_b$, being 
much narrower at NLL. In Fig.~\ref{fig2} (right) the dependence on the angle 
between the two external transverse scales, $\vec{k}_{a,b}$, is plotted. The 
region where this angle is smaller dominates and for larger energies the 
angular correlation decreases.

\subsection{Resummation of running coupling terms beyond BFKL}

Making use of the regularisation presented in this work it is 
possible to go beyond the pure BFKL calculation by resumming the 
running coupling terms. To illustrate this point it is convenient to 
connect with 
the work in Ref.~\cite{Korchemskaya:1996je} where the gluon Regge trajectory 
was calculated 
in the context of the renormalisation group evolution of Wilson lines. 
In Ref.~\cite{Korchemskaya:1996je} the trajectory is related to the so 
called ``cusp anomalous dimension'' in the form
\begin{eqnarray}
\omega_0 \left(\vec{q},\lambda\right) &=& 
- \frac{1}{2} \int_{\lambda^2}^{\vec{q}^2}
\frac{d \vec{k}^2}{\vec{k}^2} \Gamma_{\rm cusp} 
\left(\bar{\alpha}_s \left(\vec{k}^2 \right)\right) + {\rm constant},
\end{eqnarray}
with $\Gamma_{\rm cusp} \left( \bar{\alpha}_s \right) = 
\bar{\alpha}_s +  \bar{\alpha}_s^2  {\cal S}$, and 
${\cal S} = \frac{1}{3}-\frac{\pi^2}{12}+\frac{5}{12}\frac{\beta_0}{N_c}$.
One can then use, as in the original derivation of the NLL BFKL kernel, 
the ${\cal O}(\alpha_s^2)$ expansion of the coupling,
$\bar{\alpha}_s \left(\vec{k}^2\right) 
\approx \bar{\alpha}_s \left(\mu^2\right) - \bar{\alpha}_s^2 \left(\mu^2\right) \frac{\beta_0}{4 N_c}\ln{\frac{\vec{k}^2}{\mu^2}}$, to obtain the 
result of Eq.~(\ref{QCDtrajectory}) by simply fixing the constant of 
integration to the full calculation of the trajectory in 
Ref.~\cite{Fadin:1998py}, i.e., 
${\rm constant} = \bar{\alpha}_s^2 \frac{3}{2} \zeta(3)$. 

The resummation of running coupling terms is then achieved if the resummed 
running, $\bar{\alpha}_{s} \left(\vec{k}^2\right) = 
1 / \frac{\beta_0}{4 N_c}\ln{\frac{\vec{k}^2}{\Lambda_{\rm QCD}^2}}$, 
is introduced. With this choice the trajectory reads
\begin{eqnarray}
\omega_0 \left(\vec{q},\lambda \right) &=& \frac{4 N_c}{\beta_0} 
\ln{\frac{\bar{\alpha}_s \left(\vec{q}^2\right)}
{\bar{\alpha}_s \left(\lambda^2\right)}} 
+ \frac{4 N_c}{\beta_0}  {\cal S} \left[\bar{\alpha}_s \left(\vec{q}^2\right)
- \bar{\alpha}_{s} \left(\lambda^2\right)\right] +  \bar{\alpha}_s^2 \frac{3}{2} \zeta(3).
\end{eqnarray}
In this case the $\xi$ function in the real emission is 
$\xi = \bar{\alpha}_s \left(\vec{q}^2\right) 
+ \bar{\alpha}_s^2 \left(\vec{q}^2\right) \mathcal{S}$, ensuring 
the $\lambda$ independence of the GGF. This resummation will be 
further discussed in a coming publication, including a numerical study 
of the gluon--bremsstrahlung scheme, where the factor $\mathcal{S}$ is 
absorbed in the coupling constant, together with a comparison to previous 
approaches in the literature for the treatment of the running~\cite{running} 
and studies of the NLL GGF~\cite{NLLpapers}.

\section{Conformal spins in N=4 supersymmetric Yang--Mills theory}

\begin{figure}[!thb]
\vspace*{5.0cm}
\begin{center}
\includegraphics{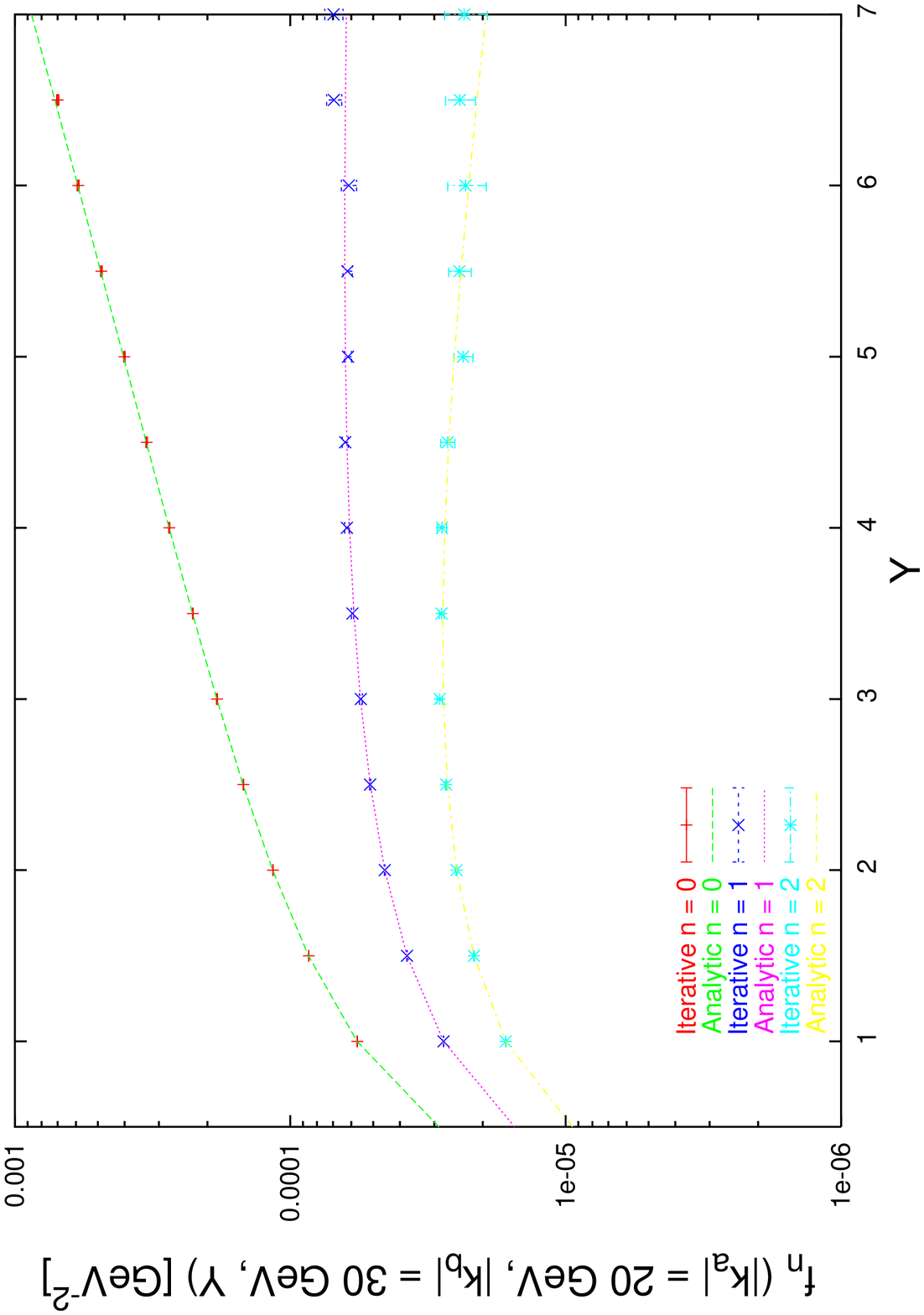}
\includegraphics{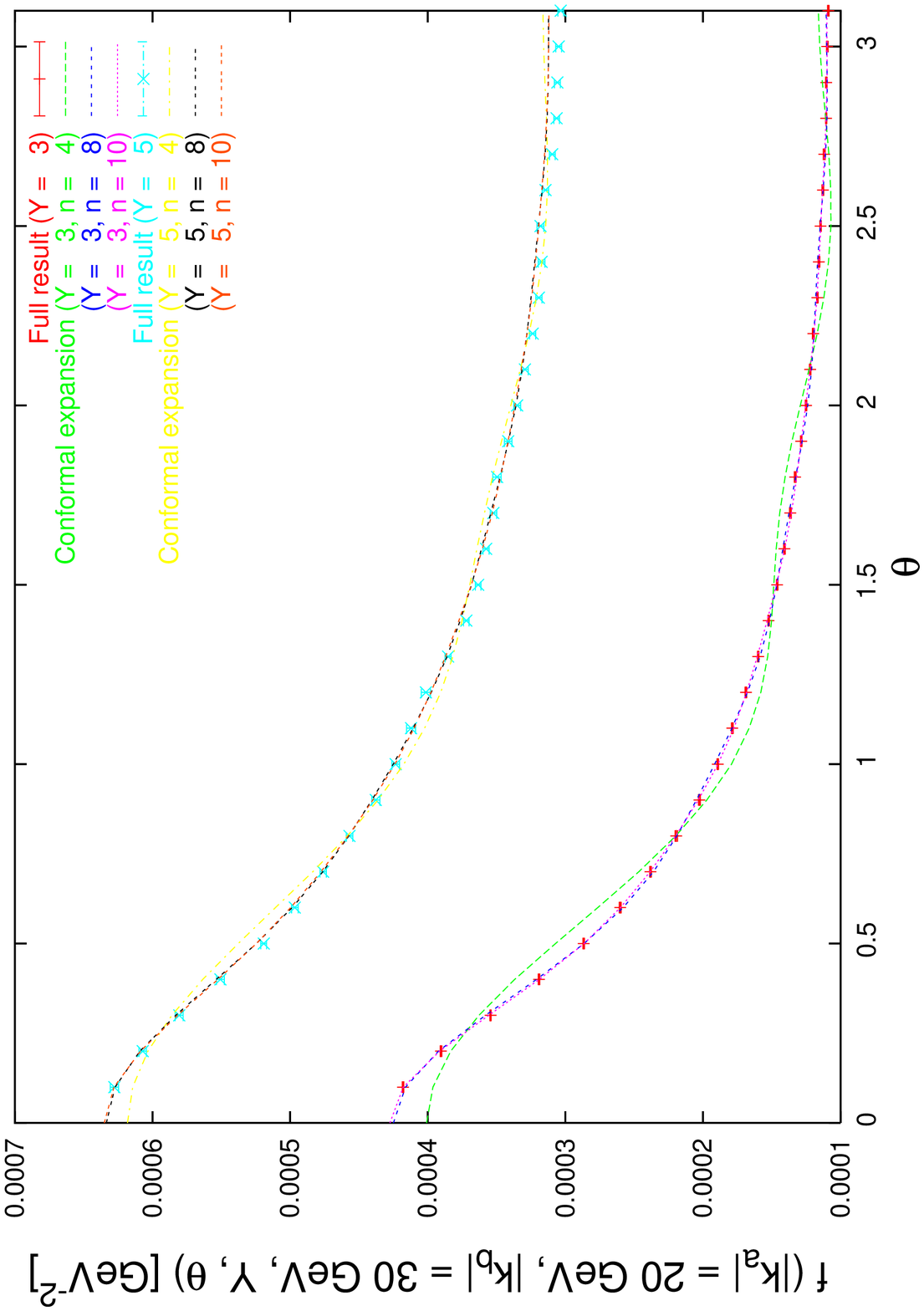}
\caption[*]{Left: Projection on conformal spins. Right: 
Convergence to the full GGF as in Eq.~(\ref{conformalexpansion}).}
\label{fig3}
\end{center}
\end{figure}

In the N=4 supersymmetric case~\cite{Kotikov} the gluon Regge trajectory in 
this regularisation simplifies to (see Ref.~\cite{Andersen:2004uj} for details)
\begin{eqnarray}
\omega_0 \left({\vec{q}},\lambda\right) &=& 
- a \left\{ \ln{\frac{\vec{q}^2}{\lambda^2}}
+ \frac{a}{4} \left[\left(\frac{1}{3}- 2 \xi(2) \right)\ln{\frac{\vec{q}^2}{\lambda^2}}- 6 \, \zeta(3) \right]\right\}, 
\label{SUSYtrajectory}
\end{eqnarray}
where now the coupling $a$ does not run, and double logarithms
 are not generated in Eq.~(\ref{SUSYtrajectory}). The expression for 
\begin{eqnarray}
\xi &\equiv& a + a^2 \left(\frac{1}{12}-\frac{\zeta(2)}{2} \right)
\end{eqnarray}
is a simple constant. As the theory is conformally invariant it is possible 
to find the solution to 
the NLL BFKL equation first expanding the GGF in conformal spins, $n$, as
\begin{eqnarray}
f\left(\vec{k}_a,\vec{k}_b, {\rm Y}\right)&=&
\sum_{n=-\infty}^{\infty} f_n\left(|\vec{k}_a|,|\vec{k}_b|, {\rm Y}\right) e^{i n \theta},
\label{conformalexpansion}
\end{eqnarray}
and then expressing the coefficients in terms of the known eigenvalues, i.e., 
\begin{eqnarray}
f_n\left(|\vec{k}_a|,|\vec{k}_b|, {\rm Y}\right) &=& 
\frac{1}{2 \pi |\vec{k}_a| |\vec{k}_b|} 
\int \frac{d \gamma}{2 \pi i} 
\left(\frac{\vec{k}_a^2}{\vec{k}_b^2}\right)^{\gamma-\frac{1}{2}}
e^{\omega_n (a,\gamma) {\rm Y}},
\label{coefficients}
\end{eqnarray}
these $\omega_n (a,\gamma)$ eigenvalues can be found 
in~\cite{Kotikov,Andersen:2004uj}. 
Alternatively, Eq.~(\ref{coefficients}) can be also obtained by means 
of the iterative method of Eq.~(\ref{TheSolution}) by projecting on the 
conformal spins:
\begin{eqnarray}
f_n\left(|\vec{k}_a|,|\vec{k}_b|, {\rm Y}\right) &=& 
\int_0^{2 \pi} \frac{d\theta}{2 \pi} \, 
f\left(\vec{k}_a,\vec{k}_b, {\rm Y}\right) \cos{\left(n \theta\right)}. 
\end{eqnarray}
The agreement found in Ref.~\cite{Andersen:2004uj} for these projections 
serves as a cross--check 
of the calculations in Ref.~\cite{Kotikov} and shows the accuracy of the 
method in Eq.~(\ref{TheSolution}). This is highlighted in Fig.~\ref{fig3} 
where the analytic expressions are compared to the iterative ones for 
the expression in Eq.~(\ref{coefficients}) (left) and the full expansion in 
Eq.~(\ref{conformalexpansion}) (right).
\vspace{-0.4cm}

\section*{Acknowledgements} 

I would like to thank J. R.~Andersen for collaboration, G.~P.~Korchemsky 
for discussions, and the 
participants of DIS 2004 for their interest in the results here presented, 
in particular: S.~Gieseke, A.~Kotikov, A.~Kyrieleis, 
L.~Lipatov, L.~Motyka, G.~Rodrigo, A.~Stasto and R.~Thorne. 
This work was supported by an Alexander von Humboldt Postdoctoral Fellowship.


\begin{thebibliography}{0}

\bibitem{FKL}
L.~N.~Lipatov,
%``Reggeization Of The Vector Meson And The Vacuum Singularity In Nonabelian Gauge Theories,''
Sov.\ J.\ Nucl.\ Phys.\  {\bf 23} (1976) 338, 
%%CITATION = SJNCA,23,338;%%    
E.~A.~Kuraev, L.~N.~Lipatov, V.~S.~Fadin,
%``The Pomeranchuk Singularity In Nonabelian Gauge Theories,''
Sov.\ Phys.\ JETP {\bf 45} (1977) 199, 
%%CITATION = SPHJA,45,199;%%
I.~I.~Balitsky, L.~N.~Lipatov,
%``The Pomeranchuk Singularity In Quantum Chromodynamics,''
Sov.\ J.\ Nucl.\ Phys.\  {\bf 28} (1978) 822.
%%CITATION = SJNCA,28,822;%%

%\cite{Fadin:1998py}
\bibitem{Fadin:1998py}
V.~S.~Fadin, L.~N.~Lipatov,
%``BFKL pomeron in the next-to-leading approximation,''
Phys.\ Lett.\ B {\bf 429} (1998) 127.
%[arXiv:hep-ph/9802290].
%%CITATION = HEP-PH 9802290;%%

%\cite{Ciafaloni:1998gs}
\bibitem{Ciafaloni:1998gs}
M.~Ciafaloni, G.~Camici,
%``Energy scale(s) and next-to-leading BFKL equation,''
Phys.\ Lett.\ B {\bf 430} (1998) 349.
%[arXiv:hep-ph/9803389].
%%CITATION = HEP-PH 9803389;%%

%\cite{Andersen:2003an}
\bibitem{Andersen:2003an}
J.~R.~Andersen, A.~Sabio Vera,
%``Solving the BFKL equation in the next--to--leading approximation,''
Phys.\ Lett.\ B {\bf 567} (2003) 116.
%[arXiv:hep-ph/0305236].
%%CITATION = HEP-PH 0305236;%%

%\cite{Andersen:2003wy}
\bibitem{Andersen:2003wy}
J.~R.~Andersen, A.~Sabio Vera,
%``The gluon Green's function in the BFKL approach at next-to-leading
%logarithmic accuracy,''
Nucl.\ Phys.\ B {\bf 679} (2004) 345.
%[arXiv:hep-ph/0309331].
%%CITATION = HEP-PH 0309331;%%

%\cite{Korchemskaya:1996je}
\bibitem{Korchemskaya:1996je}
I.~A.~Korchemskaya, G.~P.~Korchemsky,
%``Evolution equation for gluon Regge trajectory,''
Phys.\ Lett.\ B {\bf 387} (1996) 346.
%%CITATION = HEP-PH 9607229;%%

%\cite{Kotikov}
\bibitem{Kotikov}
A.~V.~Kotikov, L.~N.~Lipatov, Nucl.\ Phys.\ B {\bf 582} (2000) 19, Nucl.\ Phys.\ B {\bf 661} (2003) 19 
[Erratum-ibid.\ B {\bf 685} (2004) 405].

%\cite{Andersen:2004uj}
\bibitem{Andersen:2004uj}
J.~R.~Andersen, A.~Sabio Vera, 
%``The gluon Green's function in N = 4 supersymmetric Yang-Mills theory,''
hep-th/0406009.
%%CITATION = HEP-TH 0406009;%%

\bibitem{running} L.N. Lipatov, {JETP}  ${\bf {63}}$, 904 (1986), 
G. Camici, M. Ciafaloni,  {Phys. Lett.} B  ${\bf {395}}$, 118  (1997), 
 R.~S.~Thorne, Phys.\ Lett.\ B {\bf 474} (2000) 372, Phys.\ Rev.\ D {\bf 64} (2001) 074005, J.~R.~Forshaw, D.~A.~Ross, A.~Sabio Vera, Phys.\ Lett.\ B {\bf 498} (2001) 149, M.~Ciafaloni, D.~Colferai, G.~P.~Salam, A.~M.~Stasto, Phys.\ Lett.\ B {\bf 541} (2002) 314, Phys.\ Rev.\ D {\bf 66} (2002) 054014.

\bibitem{NLLpapers} D.A.~Ross, Phys. Lett. {\bf B431} (1998) 161, 
G.P.~Salam, JHEP{\bf 8907} (1998) 19, M.~Ciafaloni, D.~Colferai, Phys. Lett.
{\bf B452} (1999) 372, M.~Ciafaloni, D.~Colferai, G.P.~Salam,
Phys. Rev. {\bf D60} (1999) 114036, R.S.~Thorne, Phys. Rev. {\bf
D60} (1999) 054031, C.~R.~Schmidt, Phys.\ Rev.\ D {\bf 60} (1999) 074003, J.~R.~Forshaw, D.~A.~Ross, A.~Sabio Vera, Phys.\ Lett.\ B {\bf 455} (1999) 273, G.~Altarelli, R.~D.~Ball, S.~Forte, Nucl.\ Phys.\ B {\bf 575} (2000) 313, Nucl.\ Phys.\ B {\bf 621} (2002) 359, Nucl.\ Phys.\ B {\bf 674} (2003) 459, M.~Ciafaloni, D.~Colferai, G.~P.~Salam, A.~M.~Stasto, Phys.\ Lett.\ B {\bf 576} (2003) 143, Phys.\ Rev.\ D {\bf 68} (2003) 114003, Phys.\ Lett.\ B {\bf 587} (2004) 87.

\end{thebibliography}
\end{document}